%% file: main.tex
\documentclass[10pt,twocolumn,letterpaper]{article}

\usepackage{iccv}
\usepackage{times}
\usepackage{epsfig}
\usepackage{graphicx}
\usepackage{amsmath}
\usepackage{amssymb}
\usepackage{algorithm}
\usepackage{algorithmic}
\usepackage{bbm}
\usepackage{amsfonts}
\usepackage{mathrsfs}
\usepackage{multirow}
\usepackage{textcomp}
\usepackage{lipsum}
\usepackage{ulem}
\usepackage[accsupp]{axessibility}  

\usepackage[pagebackref=true,breaklinks=true,letterpaper=true,colorlinks,bookmarks=false]{hyperref}
\newcommand{\figref}[1]{\hyperref[#1]{Fig. \ref*{#1}}}
\newcommand{\tabref}[1]{\hyperref[#1]{Table \ref*{#1}}}
\newcommand{\algoref}[1]{\hyperref[#1]{Algorithm \ref*{#1}}}
\newcommand{\myeqref}[1]{\hyperref[#1]{Equation \ref*{#1}}}

\iccvfinalcopy 

\def\iccvPaperID{****} 

\begin{document}

\title{~DRNet:~ Decomposition~ and~ Reconstruction~ Network~\\for~ Remote~ Physiological~ Measurement~}
\author{Yuhang Dong,~ Gongping Yang\thanks{Corresponding author.},~  Yilong Yin\\
\vspace{-8pt}{\small~}\\
School of Software,~ Shandong University,~ China\\
{\tt\small dongyuhang42@qq.com,~ gpyang@sdu.edu.cn,~ ylyin@sdu.edu.cn}
}
\maketitle
\pagestyle{empty}  
\thispagestyle{empty}
\input{1_Abstract.tex}
\input{2_Introduction.tex}
\input{3_Related_Works.tex}
\input{4_Our_method.tex}
\input{5_Experiments.tex}
\input{7_Conclusions.tex}
\input{8_Acknowledgment.tex}
{\small
\bibliographystyle{ieee_fullname}
\bibliography{main.bib}
}
\end{document}

%% file: 1_Abstract.tex
\begin{abstract}
Remote photoplethysmography (rPPG) based physiological measurement has great application values in affective computing, non-contact health monitoring, telehealth monitoring, etc, which has become increasingly important especially during the COVID-19 pandemic. Existing methods are generally divided into two groups. The first focuses on mining the subtle blood volume pulse (BVP) signals from face videos, but seldom explicitly models the noises that dominate face video content. They are susceptible to the noises and may suffer from poor generalization ability in unseen scenarios. The second focuses on modeling noisy data directly, resulting in suboptimal performance due to the lack of regularity of these severe random noises. In this paper, we propose a Decomposition and Reconstruction Network (DRNet) focusing on the modeling of physiological features rather than noisy data. A novel cycle loss is proposed to constrain the periodicity of physiological information. Besides, a plug-and-play Spatial Attention Block (SAB) is proposed to enhance features along with the spatial location information. Furthermore, an efficient Patch Cropping (PC) augmentation strategy is proposed to synthesize augmented samples with different noise and features. Extensive experiments on different public datasets as well as the cross-database testing demonstrate the effectiveness of our approach.
\end{abstract}

%% file: 2_Introduction.tex
\section{Introduction}
Computer vision based remote physiological measurement has been gaining a tremendous interest, which has significant advantages compared with traditional contact-based measurements, particularly during the COVID-19 pandemic. Firstly, these methods can achieve reliable contactless vitals measurement such as heart rate (HR), Respiration Frequency (RF) and heart rate variability (HRV), but don't require any customized equipment. These methods rely only on the video feed recorded from a commonly accessible camera such as a commodity smartphone camera. Secondly, conventional contact-based measurements such as electrocardiography (ECG) and photoplethysmography (PPG) require dedicated skin-contact devices for data collection, which may cause discomfort and inconvenience for subjects. Hence, these computer vision based methods are more patient-friendly, which can achieve non-contact human health monitoring. Thirdly, these methods have broader applications than contact-based methods, including telehealth monitoring, deep forgery detection, affective computing, human behavior understanding and sports.

Remote photoplethysmography (rPPG) is one of the most studied computer vision based  measurement methods, which aims to extract physiological signals from video sequences. The principle of rPPG based physiological measurement is the fact that optical absorption of a local tissue varies periodically with the blood volume due to the human heartbeat. Nevertheless, the subtle optical absorption variation (not visible for human eyes) can be easily affected by the noises like head movements, lighting variations and device noises. To overcome the above challenges, a lot of conventional methods have been proposed to solve these strong random noises using color space projection~\cite{verkruysse2008remote,poh2010non,lewandowska2011measuring} or certain skin reflection models~\cite{de2013robust,wang2016algorithmic}. However, these methods are built upon the shallow and coarse assumptions, which do not always hold in handling the complicated scenes, such as large head movement or dim lighting condition. Besides, the real-world samples are usually too complex to be modeled with multiple simple mathematical models. Therefore, these methods will easily fail in real-world samples.

In recent years, deep learning has achieved significant breakthroughs in various computer vision tasks, and many scholars have also tried to utilize the strong modeling ability of deep neural networks for remote physiological signals prediction. Most of these methods focus on learning a network mapping from different manual representations of face videos (\textit{e.g.}, cropped video frames~\cite{vspetlik2018visual},  difference of video frames~\cite{chen2018deepphys}, spatial-temporal map~\cite{niu2019rhythmnet,niu2020video}). On one hand, these manual representations consist of physiological and non-physiological information, and the estimation performance can be easily affected by non-physiological information such as head movements and lighting variations. On the other hand, the key challenge of rPPG-based physiological measurement is how to effectively extract physiological information and suppress the adverse effects of non-physiological information. However, existing methods either seldom explicitly model the serious random noises, or model noisy data directly. Due to the lack of sufficient data and the lack of regularity of these strong noises, both of them lead to suboptimal performance.

Motivated by the above discussions, rather than modeling noisy data with the severe lack of regularity, it is better to directly model regular, periodic physiological features. In other words, unlike previous studies, our approach focuses on modeling physiological information rather than modeling noisy data. In this way, we propose the Decomposition and Reconstruction Network (DRNet), which adopts a novel decomposition and reconstruction strategy~(models physiological features directly, decomposes non-physiological noise indirectly, and then uses the decomposed noise and features to reconstruct new  synthetic samples). Furthermore, we propose a novel cycle loss to constrain the periodicity of physiological information.

Our contribution can be summarized as follows: 
\begin{itemize}
\item We propose the first rPPG-dedicated data augmentation method, Patch Cropping (PC), to generate data with different degrees of noise influences and multi-scale physiological information, which is able to plug and play in not only DRNet but also other existing frameworks for performance improvement.
\item To the best of our knowledge, our approach is the first to focus on modeling physiological information. A novel cycle loss is proposed to constrain the periodicity of physiological features. Moreover, a new easy-to-hard decomposition and reconstruction strategy is proposed.
\item We propose a lightweight and efficient Spatial Attention Block (SAB) plugged in our physiological estimator, which can adaptively recalibrate the varying importance of different channels and spatial regions.
\item Our proposed method achieves state-of-the-art performance on three public benchmark datasets with intra-dataset and cross-dataset testing protocols.
\end{itemize}

%% file: 3_Related_Works.tex
%
\begin{figure*}[!t]
\centering
\includegraphics[width=0.75\linewidth]{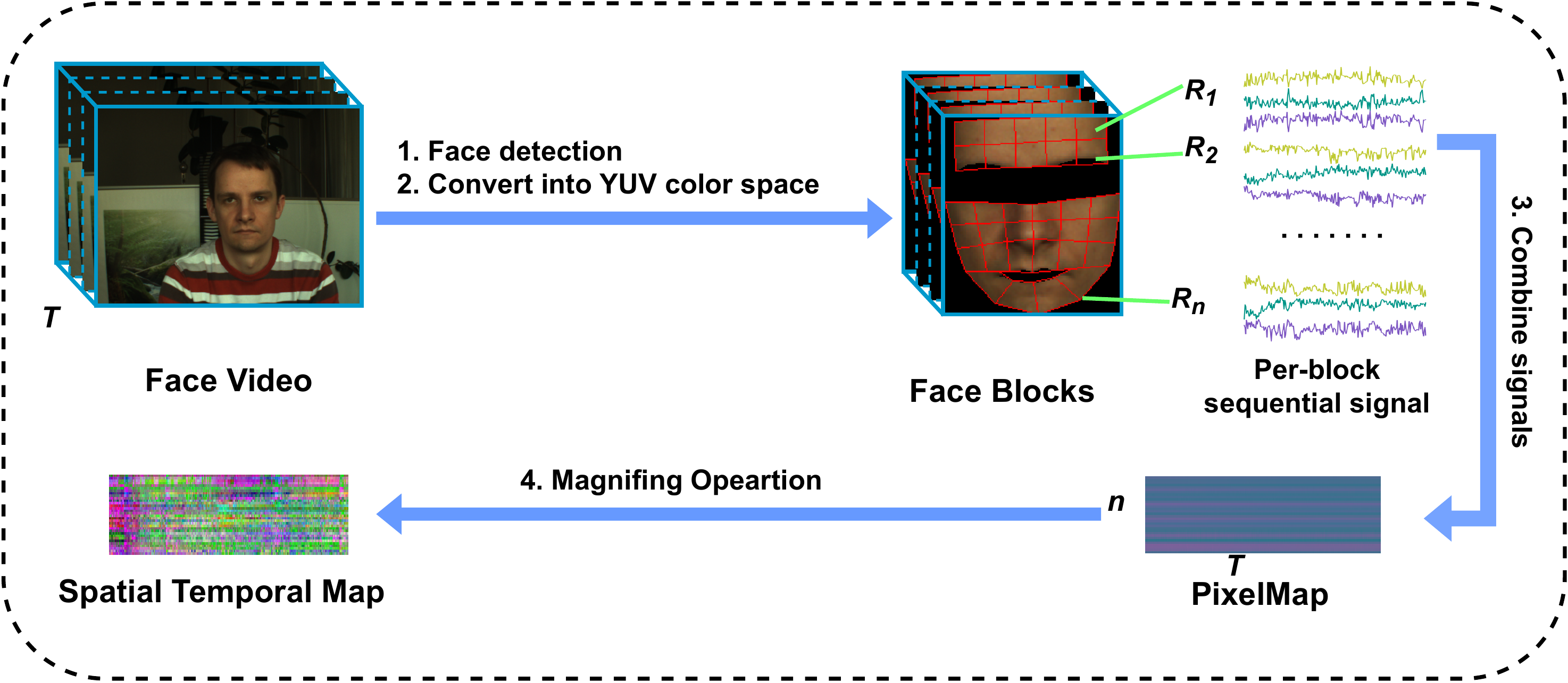}\\
\caption{Spatial Temporal Map.}
\label{fig:PC}
\end{figure*}
\section{Related Work}
\subsection{Traditional Methods}
rPPG is the monitoring of blood volume pulse (BVP) from a camera at a distance. \cite{verkruysse2008remote} proved, for the first time, that PPG signals can be measured remotely (\textgreater 1m) from face videos using ambient light. After that, many scholars have devoted their efforts in this challenging task. \cite{poh2010non} introduced a new methodology which applied independent component analysis (ICA) to reconstruct rPPG signals from face videos. Similarly, \cite{lewandowska2011measuring} proposed a new rPPG signals estimation method based on principal component analysis (PCA). In comparison to ICA, PCA can reduce computational complexity greatly. For improving motion robustness, \cite{de2013robust} proposed a CHROM method, which linearly combines the RGB channels to separate pulse signal from motion-induced distortion. Besides, \cite{wang2016algorithmic} proposed a ``plane-orthogonal-to-skin" (POS) method. Both CHROM and POS are based on the skin reflection model. These hand-crafted methods are straightforward to understand but limited in specific situations, and the estimation performance will drop significantly in complex and unconstrained scenarios.
\subsection{Deep-learning based Methods}
To overcome the limitations of traditional methods, many scholars have tried to employ deep learning (DL) technology for remote physiological measurement in recent years. The first DL-based method is DeepPhys~\cite{chen2018deepphys}, which computed the difference of frames and used an end-to-end convolutional neural network (CNN) to extract physiological signals. \cite{vspetlik2018visual} proposed the HR-CNN which predicts remote HR from aligned face images using a two-step CNN. Due to lots of irrelevant background content in raw face videos, some researchers tried to design efficient manual representations for physiological information. \cite{niu2019rhythmnet} designed a novel and efficient spatial-temporal map, which is mapped by a CNN to its HR value, and used a CNN-RNN structure to predict average HR values. In addition, \cite{niu2020video} attempted to remove the noise via cross-verified feature disentangling. \cite{lu2021dual} tried to use Dual-GAN to model noise distribution and physiological estimator directly. \cite{yu2021physformer} proposes an end-to-end video transformer based architecture, to adaptively aggregate both local and global spatio-temporal features for rPPG representation enhancement. These methods using deep networks can be divided into two categories. Methods in the first group~(\textit{e.g.}, \cite{chen2018deepphys,niu2019rhythmnet,yu2021physformer}), focus on mining the subtle BVP signals from face videos, but seldom explicitly model the noises that dominate face video content. They are susceptible to the noises and may suffer from poor generalization ability in unseen scenarios. The second category of methods, which are more close to our approach, focus on directly disentangling the physiological information with non-physiological representations~\cite{niu2020video,lu2021dual}. However, because of the lack of regularity of these serious random noises, these methods are difficult to converge, which leads to suboptimal performance.

%% file: 4_Our_method.tex
\section{Methodology}
We denote the input video as $v$, and the corresponding ground-truth BVP signal as $s_{gt}$. The goal of remote physiological measurement is to learn a mapping:
\begin{equation}
F: v \rightarrow s_{gt}
\end{equation}
\subsection{Spatial Temporal Map}
\begin{algorithm}[t]
\small
\caption{Patch Cropping Augmentation}
\label{alg:algorithm}
\textbf{Input}: Original STMap $m$; Enlarged STMap $m_{e}$; Cropping probability $\rho$, Original STMap size $C \times n \times T$; Enlarged STMap size $C \times n_{e} \times T$.
\begin{algorithmic}[1] 
\STATE $\rho_{1} \leftarrow Rand(0, 1)$.
\IF {$\rho_{1} \ge \rho$}
\STATE $m_{c} \leftarrow m$;
\ELSE
\STATE $x \leftarrow 0, y \leftarrow Rand(0, n_{e} - n)$;
\STATE $W \leftarrow T, H \leftarrow n$
\STATE $area \leftarrow (x, y, x + W, y + H)$;
\STATE $m_{c} \leftarrow m_{e}(area)$;
\ENDIF
\STATE \textbf{return} Cropped STMap $m_{c}$. 
\end{algorithmic}
\end{algorithm}
\label{section:sec3.1}
Many previous methods~\cite{vspetlik2018visual,tsou2020siamese} focused on direct applying CNNs to the face videos with good results. However, due to the low PSNR of rPPG signals in face videos, these methods are inefficient, expensive and time-consuming. Hence, in order to avoid high computational complexity and time-consuming, we choose to use Spatial Temporal Map (STMap) like \cite{niu2019rhythmnet}, which establishes a preliminary representation of the physiological signal by discarding most of the irrelevant background content.

An illustration of STMap generation procedure can be found in \figref{fig:PC}. Firstly, a face detector is applied to the face video to obtain the face position and 68 facial landmarks. Secondly, the face is cropped from each frame of the video and converted to YUV color space. Thirdly, the face images is divided into $n$ grids according to the predefined ROI. The average of the pixel values of each grid is calculated and then concatenated into a sequence of $T$ for $C$ channels. The $n$ grids are directly placed into rows. We denote the combined signal sequences as PixelMap. Finally, Magnifying Operation, a max-min normalization, is applied to all the temporal sequences of the PixelMap to scale the temporal series into $[0, 255]$. The PixelMap and STMap have the same dimension $C \times n \times T$, in which $n$ denotes the number of ROIs, $T$ denotes the number of frames of a video clip, and $C = 3$ denotes the three channels of R, G and B. 

We denote the PixelMap and STMap computed from $v$ as $pm$ and $m$, respectively. Then our goal is to establish a mapping:
\begin{equation}
F_{b}: m \rightarrow s_{gt}
\end{equation}
Unlike the previous design of STMap~\cite{niu2019rhythmnet}, we have made the following improvements: (1) We detect faces using RetinaFace~\cite{deng2020retinaface} with MobileNet~\cite{howard2017mobilenets} backbone, which can get more precise face landmarks faster. (2) We appropriately reduced the ROI area by discarding non-skin facial areas such as eyes and mouth region. An example of our ROI definition consisting of $n = 32$ ROIs is shown in the \figref{fig:PC}.
%
\begin{figure*}[!t]
\centering
\includegraphics[width=0.75\linewidth]{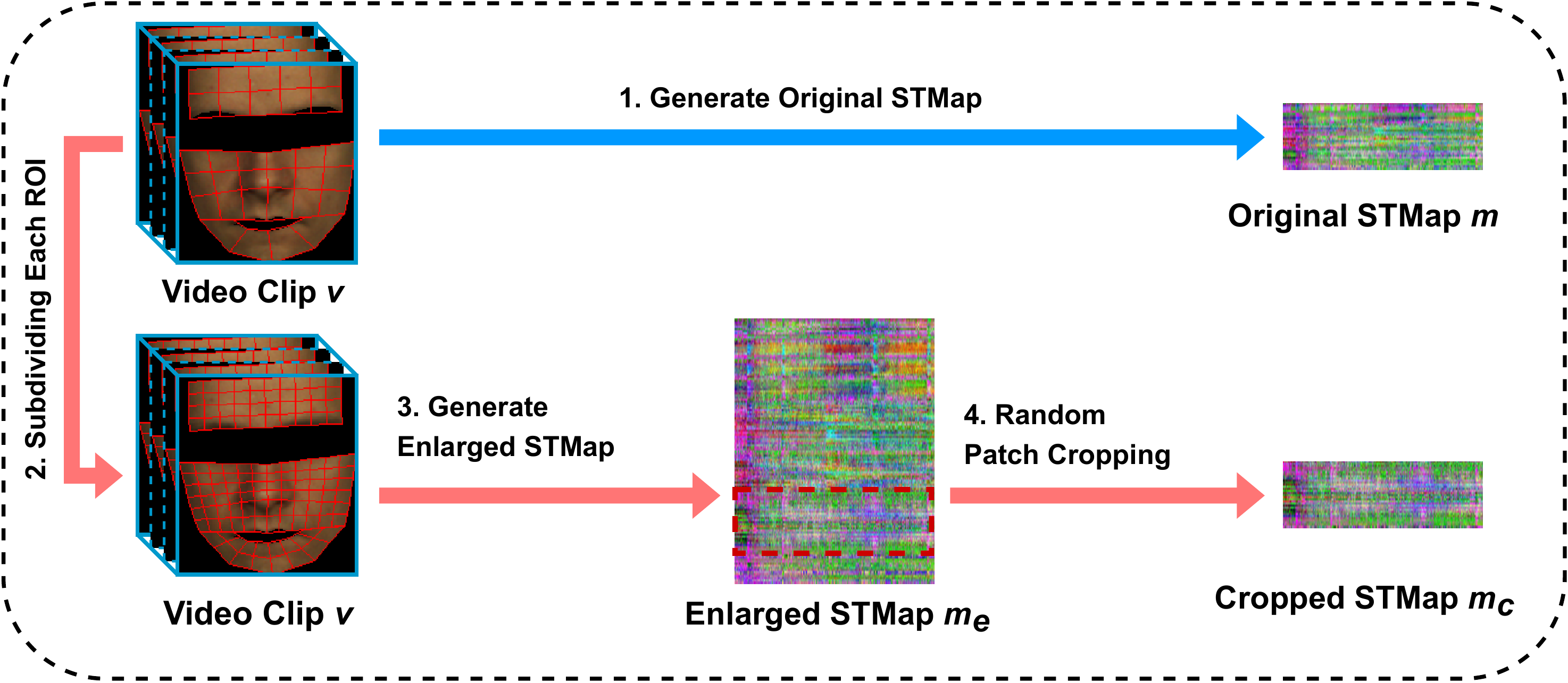}\\
\caption{Patch Cropping Augmentation.}
\label{fig:PC1}
\end{figure*}
\subsection{Patch Cropping Augmentation}
Due to the high collection cost for remote physiological measurement, there are limited data scale and diversity in public datasets. Besides, since the amplitude of the optical absorption color change is very small, applying traditional augmentation methods (\textit{e.g.}, random cropping or flipping) directly may destroy the subtle physiological information. To address these issues, we propose an rPPG-dedicated data augmentation method, named Patch Cropping (PC), to synthesize new samples with different noise levels. 

The illustration and algorithm of PC are summarized in \figref{fig:PC1} and \algoref{alg:algorithm} respectively. An enlarged STMap $m_{e}$ can be generated by subdividing each face ROI into $\gamma \times \gamma$ sub-ROIs. Besides, the enlarged STMap $m_{e}$ has the dimension of $C \times n_{e} \times T$, in which $n_{e} = n \times \gamma \times \gamma$. Specifically, a total of $32$ face ROIs are defined in the process of generating the original STMap $m$. When $\gamma$ is set to $2$, each face ROI is divided into $2 \times 2$ sub-ROIs to get a new ROI definition consisting of 128 ROIs, in which $128 = 32 \times 2 \times 2$. After that, these two ROI definitions are used to generate the original STMap $m$ of dimension $3 \times 32 \times 256$ and the enlarged STMap $m_{e}$ of dimension $3 \times 128 \times 256$, respectively. Finally, without direct interpolating, lossless data augmentation is achieved by randomly cropping the enlarged STMap $m_{e}$. The two hyperparameters $\gamma$ and $\rho$ control the enlared ratio and the augmentation intensity, respectively. As a tradeoff, we use empirical settings $\gamma = 2$ and $\rho = 0.5$ for experiments.

There are three advantages for PC: (1) In PC, the cropped STMap $m_{c}$ and the original STMap $m$ describe the subject’s physiological information from detailed and general aspects respectively. (2) In PC, a more detailed STMap $m_{e}$ is obtained by subdividing each face ROI, which avoids the computational error of other existing methods (\textit{e.g.}, random cropping, upsampling and downsampling~\cite{niu2019robust}) caused by direct cropping or interpolating. (3) Compared with the original STMap $m$, the cropped STMap $m_{c}$ contains similar and multi-scale physiological information and different levels of noise influences, not achieved by other existing methods (\textit{e.g.}, random horizontal and vertical flipping~\cite{niu2020video}).
\subsection{DRNet}
\begin{figure*}[!t]
\centering
\includegraphics[width=0.9\linewidth]{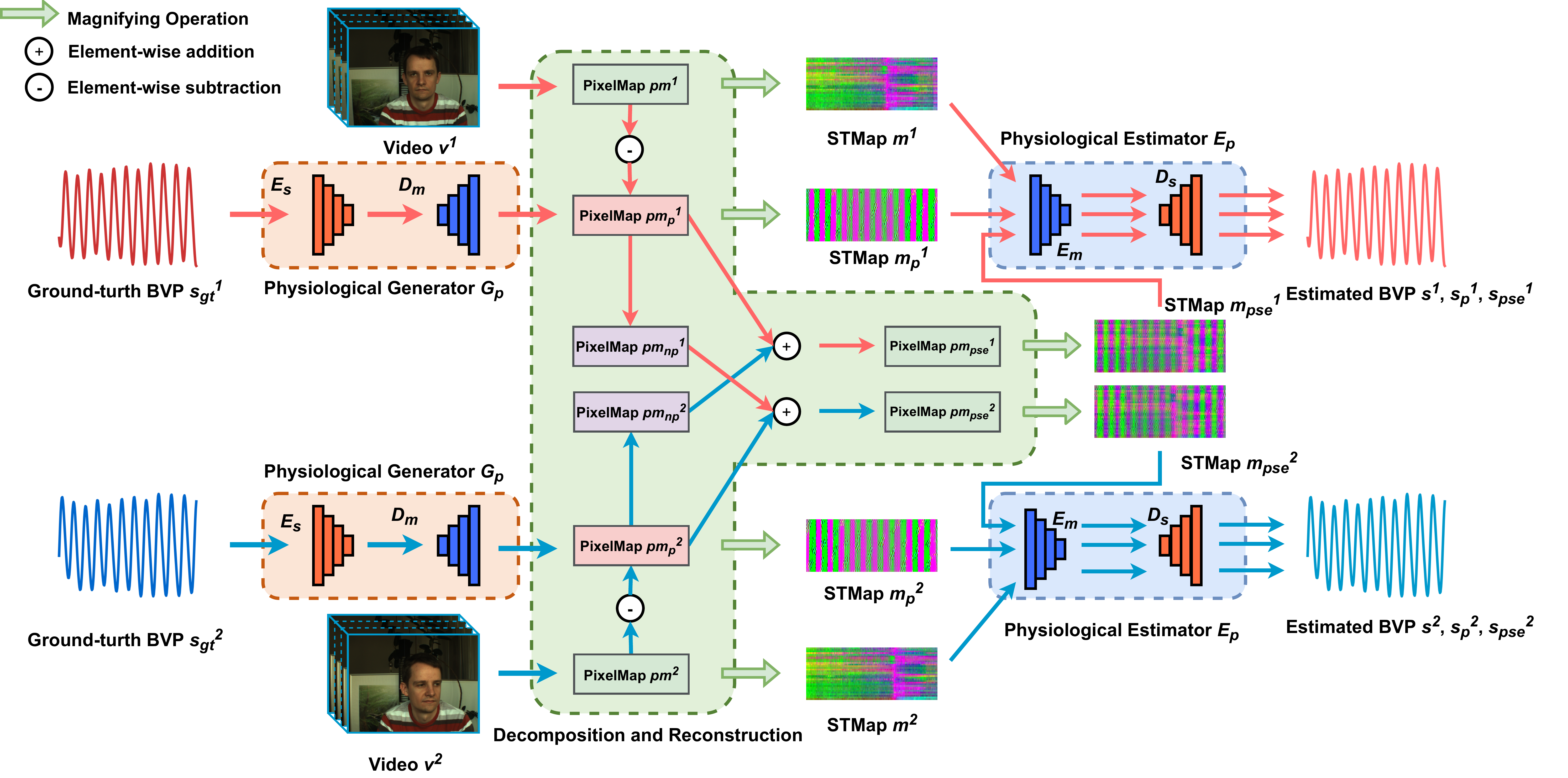}\\
\caption{Framework of the DRNet. Pairwise face video chips are used for training. We first generate the corresponding PixelMaps $pm^{1}$, $pm^{2}$ of the input face video clips. Then we feed the ground-truth BVP signals into the Physiological Generator $G_{p}$ to generate the noise-free PixelMaps. After that, we cross-generate the pseudo PixelMaps $pm_{pse}^{1}$, $pm_{pse}^{2}$ using different combinations of noises and features from different original PixelMaps.  The physiological estimator $E_{p}$ takes the STMaps which is generated by the ground-truth PixelMaps $pm^{1}$, $pm^{2}$ and the synthetic PixelMaps $pm_{p}^{1}$, $pm_{p}^{2}$, $pm_{pse}^{1}$, $pm_{pse}^{2}$ through magnifying operation as input for rPPG signal predictions. The modules of the same type in our network use shared weights.}
\label{fig:pipeline}
\end{figure*}
\paragraph{Overall Architecture} In our task, noisy data consists of physiological features and non-physiological noise. Specifically, the creation of a PixelMap $pm$ can be generally formulated by a linear model:
\begin{equation}
\label{eq:pm_model}
pm = pm_{p} + pm_{np}
\end{equation}
where $pm_{p}$ represents the physiological information which is a noise-free, regular, periodic optical absorption color variation purely caused by the rPPG signal $s_{gt}$, and $pm_{np}$ denotes the non-physiological information which is the summation of all noise sources physically caused by external environment such as head movements and device noises. Existing methods either seldom explicitly model the noise $pm_{np}$, or model the noisy data $pm$ or the noise $pm_{np}$ directly. Due to the lack of sufficient data and the lack of regularity of these strong noises, both of them lead to suboptimal performance.

To address these problems, DRNet follows the strategy of decomposition and reconstruction. To achieve indirect feature disentangling, we focus on modeling the noise-free and periodic physiological information $pm_{p}$, in which its generation method is fixed, regular and easy to fit, and then decompose the noise $pm_{np}$ from the noisy data $pm$ using \myeqref{eq:pm_model}. After that, we cross-generate the synthetic data using the decomposed features and noise for reconstructing new samples.

Specifically, as shown in \figref{fig:pipeline}, with pairwise input face video clips $v^{1}$, $v^{2}$ and the  corresponding ground-truth BVP signals $s_{gt}^{1}$, $s_{gt}^{2}$, we first generate the corresponding PixelMaps $pm^{1}$ and $pm^{2}$ computed from $v^{1}$, $v^{2}$ respectively. Then, we use the Physiological Generator $G_{p}$ to generate the PixelMaps $pm_{p}^{1}$, $pm_{p}^{2}$ from the ground-truth BVP signals $s_{gt}^{1}$, $s_{gt}^{2}$ respectively.

After that, we apply subtraction operation of subtracting $pm_{p}^{1}$, $pm_{p}^{2}$ from $pm^{1}$, $pm^{2}$ to separate out the non-physiological information $pm_{np}^{1}$, $pm_{np}^{2}$ respectively. Moreover,  pseudo PixelMaps $pm_{pse}^{1}$, $pm_{pse}^{2}$ are generated by using different combinations of noises and features, \textit{i.e.}, pseudo PixelMap $pm_{pse}^{1}$ is generated by adding $pm_{p}^{1}$ to $pm_{np}^{2}$, and $pm_{pse}^{2}$ is generated by adding $pm_{p}^{2}$ to $pm_{np}^{1}$.

Finally, the STMaps which is generated by the ground-truth PixelMaps $pm^{1}$, $pm^{2}$ and the synthetic PixelMaps $pm_{p}^{1}$, $pm_{p}^{2}$, $pm_{pse}^{1}$, $pm_{pse}^{2}$ through Magnifying Operation are both fed to the physiological estimator $E_{p}$ for physiological signal predictions.
\paragraph{Loss function} For rPPG signal prediction, we use a Pearson correlation based loss to define the similarity between the predicted signal and ground truth. Specifically, the loss function is given by
\begin{equation}
\mathcal{L}_{phy} = 1 -\frac{\sum_{t = 1}^{T}( x^{i} - \overline{x} )( y^{i} - \overline{y} ) }{\sqrt{\sum_{t = 1}^{T}(x^{i} - \overline{x})^2 } \sqrt{\sum_{t = 1}^{T}(y^{i} - \overline{y})^2} }
\end{equation}
where $x$ is the ground-truth rPPG signals, $y$ is the estimated rPPG signals, $\overline{x}$ and $\overline{y}$ denote the mean values of $x$ and $y$.

In addition, we propose a cycle loss to constrain the periodicity of the generated PixelMap $pm_{p}$ which represents periodic optical absorption color change. The cycle loss averages randomly selected $n_{c}=Rand(1, n)$ rows from $pm_{p}$. We denote the averaged result as $pm_{avg}$ with dimension of $1 \times T \times c$. Finally, we constrain the periodicity of $c$ channels of $pm_{avg}$ respectively.
\begin{equation}
\mathcal{L}_{cyc} = \frac{1}{c}\sum_{i = 1}^{c}CE(PSD(pm_{avg}^{i}), HR_{gt})
\end{equation}
where $pm_{avg}^{i}$ denotes the $i$-th channel of $pm_{avg}$, $HR_{gt}$ indicates the ground-truth HR, $PSD(\cdot)$ indicates the power spectral density of $pm_{avg}^{i}$, and $CE(\cdot)$ indicates the cross-entropy loss. The overall loss function of our DRNet is
\begin{equation}
\mathcal{L} = \mathcal{L}_{phy} + \mathcal{L}_{cyc}
\end{equation}
\subsection{Spatial Attention Block}
\begin{figure*}[!t]
	\centering
	\includegraphics[width=0.9\linewidth]{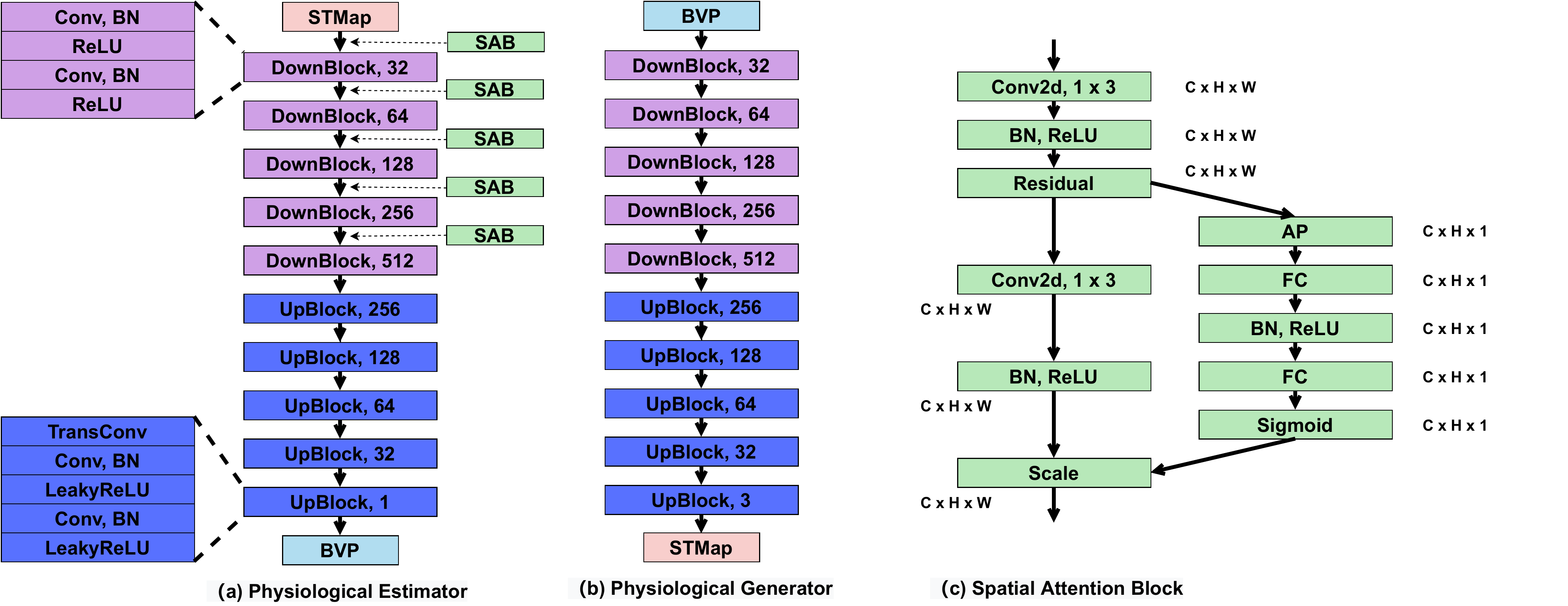}\\
	\caption{The architecture of the (a) physiological estimator $E_{p}$, (b) physiological generator $G_{p}$, (c) spatial attention block. ``BN'' denotes one batch normalization layer, ``AP'' denotes averaging pooling, ``FC'' denotes one linear layer.}
	\label{fig:SAB}
\end{figure*}
For remote physiological signal estimation, what the network really needs to focus on is the faint color variation caused by rPPG. However, noise such as head movements, non-skin regions, occlusion, illumination changes can destroy precise rPPG signal predictions. One idea is to use heavy preprocessing such as skin segmentation or ROI selection algorithms on face videos to reduce the impact of noise. However, this solution is inefficient and time-consuming. In summary, it is very important to use attention mechanism to effectively extract physiological features. Therefore, we introduce a plug-and-play Spatial Attention Block (SAB) to efficiently utilize physiological information using lightweight computation and memory.

Unlike previous methods~(\textit{e.g.}, \cite{lu2021dual}) using highly redundant attention mechanisms, SAB contains very few parameters. Each row of STMap $m$ represents the raw temporal signal for one ROI on the face, and the input STMap is split by rows. Therefore, as shown in \figref{fig:SAB}, 2D-Conv with kernel size $1 \times 3$ is performed. Inspired by SENet~\cite{hu2018squeeze}, our attention module is added to the input feature map in both spatial and channel dimensions by adaptively recalibrating the varying importance of different channels and spatial face ROIs. The proposed SAB is lightweight and efficient, which can help our model exploit more informative features.

%% file: 5_Experiments.tex
\section{Experiments}
\subsection{Datasets}
In this section, we evaluate the effectiveness of the proposed method on three public-domain datasets.
\paragraph{VIPL-HR} VIPL-HR dataset~\cite{dbVIPLHR} is a challenging large-scale multi-modal database, which contains 2,378 visible light facial videos of 107 subjects. In order to simulate real world conditions as realistic as possible, this dataset was collected under less-constrained scenarios, which contains various variations such as different head movements, illumination condition variations, and acquisition device changes.
\paragraph{PURE} PURE dataset~\cite{dbPURE} is a public available database for remote heart rate estimation, which comprises 60 RGB videos from 10 subjects(8 male, 2 female) in 6 different setups. The ground-truth rPPG signals were captured using a finger clip pulse oximeter (pulox CMS50E).
\paragraph{UBFC-rPPG} UBFC-rPPG dataset~\cite{dbUBFCrPPG} is a database for remote heart rate estimation, which contains 42 uncompressed RGB videos. In order to make this dataset cover a wider range of heart rate values, all subjects were asked to play a time sensitive mathematical game that supposedly raises their heart rate.
\subsection{Evaluation Metrics}
We perform HR estimation on VIPL-HR, PURE and UBFC-rPPG, and cross-database HR estimation on UBFC-rPPG with training on PURE. For HR estimation, we follow \cite{niu2020video}, and standard deviation of the error (Std), mean absolute error (MAE), root mean square error (RMSE), mean error rate percentage (MER), and Pearson’s correlation coefficient (r) are employed for performance evaluation.
\subsection{Implementation Details}
Firstly, our pipeline was implemented using PyTorch framework and trained on one NVIDIA GeForce RTX 3090 GPU. We train the DRNet for 40 epochs, using random initialization. Adam optimizer~\cite{kingma2014adam} is used while learning rate is set to 0.0001 and batch size is set to 32. Secondly, in all experiments, the length of each video clip $T$ is set to 256 frames, and the step between clips is 10 frames. Besides, we pre-processed the ground truth rPPG signal using a 4th-order Butterworth bandpass filter with cutoff frequency $[0.6, 3]$ Hz for restricting outliers like \cite{wang2018single}. We follow the previous studies~\cite{tsou2020siamese,niu2020video,lu2021dual} to compute HR. Finally, all face videos and the corresponding rPPG signals were resampled to 30 fps using cubic spline interpolation like \cite{lu2021dual} before generating STMap. In order to simplify the training process, we pre-train $E_{s}$ and $D_{s}$ as a rPPG signal autoencoder on two public PPG datasets (BIDMC~\cite{pimentel2016toward} and Cuff-less~\cite{kachuee2015cuff}), and fix their weights when training the DRNet.
\subsection{Results}
%
%
\begin{table}[t]
\caption{HR estimation results by our method and several state-of-the-art methods on the VIPL-HR database.}
\small
\centering
\setlength\tabcolsep{3pt}
\begin{tabular}{llllll}
\hline
Method & Reference & Std$\downarrow$ & MAE$\downarrow$ & RMSE$\downarrow$ & r$\uparrow$ \\
\hline
SAMC~\cite{tulyakov2016self}         & CVPR   & 18.0  & 15.9 & 21.0 & 0.11     \\
POS~\cite{wang2016algorithmic}       & TBME   & 15.3  & 11.5 & 17.2 & 0.30     \\
CHROM~\cite{de2013robust}            & TBME   & 15.1  & 11.4 & 16.9 & 0.28     \\
I3D~\cite{carreira2017quo}           & CVPR   & 15.9  & 12.0 & 15.9 & 0.07     \\
DeepPhys~\cite{chen2018deepphys}      & ECCV   & 13.6  & 11.0 & 13.8 & 0.11     \\
PhysNet~\cite{yu2019remote}          & BMVC   & 14.9  & 10.8 & 14.8 & 0.20     \\
AutoHR~\cite{yu2020autohr}           & SPL    & 8.48  & 5.68 & 8.68 & 0.72     \\
ST-attention~\cite{niu2019robust}    & FG     & 7.99  & 5.40 & 7.99 & 0.66     \\
RhythmNet~\cite{niu2019rhythmnet}    & TIP    & 8.11  & 5.30 & 8.14 & 0.76     \\
NAS-HR~\cite{lu2021hr}               & VRIH   & 8.10  & 5.12 & 8.01 & 0.79     \\
CVD~\cite{niu2020video}              & ECCV   & 7.92  & 5.02 & 7.97 & 0.79     \\
PhysFormer~\cite{yu2021physformer}   & CVPR   & 7.74  & 4.97 & 7.79 & 0.78     \\
Dual-GAN~\cite{lu2021dual}           & CVPR   & 7.63  & 4.93 & 7.68 & 0.81     \\
\textbf{DRNet(Ours)}                & -      & \textbf{6.75}  & \textbf{4.18} & \textbf{6.78} & \textbf{0.85}     \\
\hline
\end{tabular}
\label{tab:vipl_hr_result}
\end{table}
%
%
%
%
\begin{table}[t]
\small
\centering
\setlength\tabcolsep{4pt}
\caption{HR estimation results by our method and several state-of-the-art methods on the PURE database.}
\begin{tabular}{llllll}
\hline
Method & Reference & MAE$\downarrow$ & RMSE$\downarrow$ & r$\uparrow$ \\
\hline
2SR~\cite{de2014improved}            & Physiol Meas & 2.44 & 3.06 & 0.98 \\
CHROM~\cite{de2013robust}            & TBME         & 2.07 & 9.92 & 0.99 \\
HR-CNN~\cite{vspetlik2018visual}     & BMVC         & 1.84 & 2.37 & 0.98 \\
Dual-GAN~\cite{lu2021dual}           & CVPR         & 0.82 & 1.31 & 0.99 \\
\textbf{DRNet(Ours)}                & -            & \textbf{0.39}  & \textbf{0.52} & \textbf{0.998} \\
\hline
\end{tabular}
\label{tab:pure_hr_result}
\end{table}
%
%
\paragraph{HR estimation on VIPL-HR} Following the original protocol in \cite{niu2019rhythmnet} for a fair comparison, a subject-exclusive 5-fold cross-validation protocol is used on the large-scale VIPL-HR dataset. We compare our method with several baseline methods on VIPL-HR, in which the performance of these baseline methods are directly from \cite{yu2021physformer}. As illustrated in \tabref{tab:vipl_hr_result}, all three traditional methods~(SAMC~\cite{tulyakov2016self}, POS~\cite{wang2016algorithmic} and CHROM~\cite{de2013robust}) perform quite poorly, struggling to handle the unconstrained complex scenarios~(\textit{e.g.}, large head movement and various lighting condition). Similarly, the end-to-end methods~(\textit{e.g.}, DeepPhys~\cite{chen2018deepphys} and PhysNet~\cite{yu2019remote}) using raw face videos as input also perform poorly due to the lack of efficient data representation. Directly processing raw facial video is not only time-consuming but also computationally intensive. Moreover, its low signal-to-noise ratio is not conducive to the extraction of rPPG signals. The proposed approach is far outperforms all the remaining methods~(\textit{e.g.}, RhythmNet~\cite{niu2019rhythmnet} and Dual-GAN~\cite{lu2021dual}) under all measures because DRNet incorporates a novel and powerful data augmentation strategy and an efficient attention module. Such results also demonstrate the superiority of our approach, which focuses not on noisy data modeling like previous methods but on physiological information modeling.

\paragraph{HR estimation on PURE and UBFC-rPPG} We further evaluate the effectiveness of the proposed approach by performing HR estimation on two small datasets (PURE and UBFC-rPPG). For PURE dataset, we follow the same testing protocol in \cite{vspetlik2018visual} for a fair comparison. The performance of 2SR, CHROM, HR-CNN, Dual-GAN are available in \cite{lu2021dual}. As shown in \tabref{tab:pure_hr_result}, the proposed DRNet outperforms all the baseline methods under all measures. For UBFC-rPPG dataset, we follow the same testing protocol in \cite{lu2021dual}, in which the videos of the first 30 subjects are used for training, and the videos of the remaining 12 subjects are used for testing. The results of POS, CHROM, GREEN, SysRhythm, Dual-GAN are directly from \cite{lu2021dual}, and the results of PulseGAN are directly from \cite{song2021pulsegan}. As illustrated in \tabref{tab:ubfc_hr_result}, our method still outperforms the baseline methods. These results show that our method can not only perform well on large-scale datasets, but also on small datasets.
%
%
\begin{table}[t]
\small
\centering
\setlength\tabcolsep{2pt}
\caption{HR estimation results by our method and several state-of-the-art methods on the UBFC-rPPG database.}
\begin{tabular}{llllll}
\hline
Method & Reference & MAE$\downarrow$ & RMSE$\downarrow$ & MER$\downarrow$ & r$\uparrow$ \\
\hline
POS~\cite{wang2016algorithmic}       & TBME          & 8.35 & 10.00 & 9.85\% & 0.24 \\
CHROM~\cite{de2013robust}            & TBME          & 8.20 & 9.92  & 9.17\% & 0.27 \\
GREEN~\cite{verkruysse2008remote}    & Opt. Express  & 6.01 & 7.87  & 6.48\% & 0.29 \\
SynRhythm~\cite{niu2018synrhythm}    & ICPR          & 5.59 & 6.82  &  5.5\% & 0.72 \\
PulseGAN~\cite{song2021pulsegan}     & JBHI          & 1.19 & 2.10  & 1.24\% & 0.98 \\
Dual-GAN~\cite{lu2021dual}           & CVPR          & 0.44 & 0.67  & \textbf{0.42\%} & 0.99 \\
\textbf{DRNet(Ours)}                & -             & \textbf{0.42}  & \textbf{0.64} & 0.45\% & \textbf{0.998}\\
\hline
\end{tabular}
\label{tab:ubfc_hr_result}
\end{table}
%
%
\begin{table}[t]
\small
\centering
\setlength\tabcolsep{2pt}
\caption{Cross-database HR estimation (training on PURE and testing on UBFC-rPPG) by our DRNet and baseline methods.}
\begin{tabular}{llllll}
\hline
Method & Reference & MAE$\downarrow$ & RMSE$\downarrow$ & MER$\downarrow$ & r$\uparrow$ \\
\hline
GREEN~\cite{verkruysse2008remote}    & Opt. Express  & 8.29 & 15.82 & 7.81\% & 0.68 \\
ICA~\cite{poh2010non}                & Opt. Express  & 4.39 & 11.60 & 4.30\% & 0.82 \\
POS~\cite{wang2016algorithmic}       & TBME          & 3.52 &  8.38 & 3.36\% & 0.90 \\
CHROM~\cite{de2013robust}            & TBME          & 3.10 &  6.84 & 3.83\% & 0.93 \\
PulseGAN~\cite{song2021pulsegan}     & JBHI          & 2.09 &  4.42 & 2.23\% & 0.97 \\
Siamese-rPPG~\cite{tsou2020siamese}  & ACM SAC       & 1.29 &  8.73 & -      & -    \\
Dual-GAN~\cite{lu2021dual}           & CVPR          & 0.74 &  \textbf{1.02} & 0.73\% & 0.997 \\
\textbf{DRNet(Ours)}                & -             & \textbf{0.65}  & 1.29 & \textbf{0.68\%} & \textbf{0.997}\\
\hline
\end{tabular}
\label{tab:pure_ubfc_hr_result}
\end{table}
\paragraph{Cross-database HR estimation} Cross-dataset evaluations are conducted to testify the generalization ability of our solution under unseen scenarios. We follow PulseGAN~\cite{song2021pulsegan} and Dual-GAN~\cite{lu2021dual} to train our model on PURE and valid it on UBFC-rPPG. As shown in \tabref{tab:pure_ubfc_hr_result}, the results of GREEN, ICA, POS, CHROM, PulseGAN are from \cite{song2021pulsegan}. Besides, the results of Siamese-rPPG and Dual-GAN are from \cite{tsou2020siamese} and \cite{lu2021dual} respectively. It can be seen from \tabref{tab:pure_ubfc_hr_result} that the DRNet is still very effective for HR estimation under unseen scenarios. These results indicate that the proposed method has a better generalization ability under new scenarios with unknown noises.
\subsection{Ablation Study}
In this subsection, all ablation studies are conducted for the proposed method for HR estimation on the large-scale and challenging VIPL-HR database. We train our network using just the physiological estimator $E_{p}$ and the ground-truth STMaps as the baseline method.
\paragraph{Effectiveness of PC} As shown in \tabref{tab:ablation_vipl_hr}, PC can improve the performance of HR estimation on VIPL-HR dataset greatly. The vast majority of previous studies lack effective data augmentation strategies. And compared with other existing augmentation methods, PC for the first time realizes the utilization of multi-scale features without destroying the very weak physiological information. It is worth noting that the augmented samples contain different degrees of noise influences and multi-scale physiological information, which is why PC can make remarkable performance gains.
\paragraph{Effectiveness of SAB} It can be seen from the second and third rows of \tabref{tab:ablation_vipl_hr} that the baseline method without SAB achieves worse when without enhancing features along with the spatial location information. In contrast, compared with the baseline method without SAB, the baseline method assembled with SAB obtains 5.09\% MAE decrease, which indicates the effectiveness of SAB. Video processing is a time-sensitive task, and the low-parameter SAB can effectively and efficiently extract physiological signals by adaptively adjusting the weights of different channels and face ROIs.
\paragraph{Effectiveness of DRNet} 
\begin{figure}[t]
	\centering
	\includegraphics[width=\linewidth]{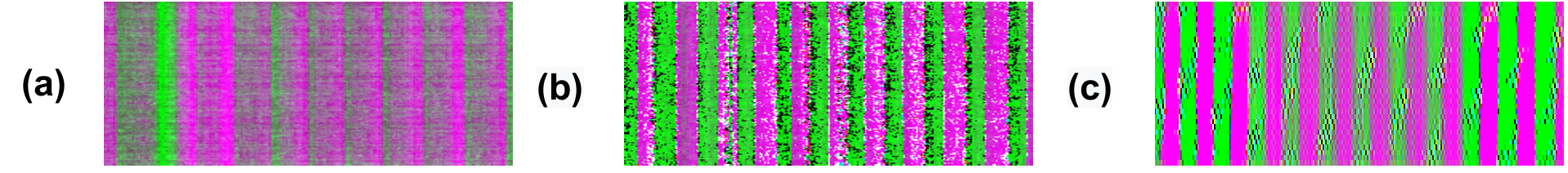}\\
	\caption{(a) The STMap with real and small noises computed from facial video that is collected in constrained condition. (b) The noise-less STMap generated by detrending all the temporal sequences of the STMap (a). (c) The synthetic noise-free STMap generated by our method.}
	\label{fig:Result_Cmp}
\end{figure}
As illustrated in \tabref{tab:ablation_vipl_hr}, with the help of SAB and PC, DRNet makes the prediction results further improved and can reduce the MAE and RMSE errors by 0.29 and 0.51. In addition, as shown in \figref{fig:Result_Cmp}, we visualize the synthetic noise-free STMap generated by our method and the real noise-less STMap, and we can see that the synthetic noise-free STMap is very similar to the real noise-less STMap. This shows that our network can decompose and model physiological information well, which is a problem not well addressed by previous methods.

The advantages of DRNet are three-fold: (1)  Generating training examples for supervised tasks is a long sought after goal in AI, and DRNet presents a novel data synthesis strategy without modeling noise or noisy data directly. (2) DRNet separates out non-physiological noise by explicitly modeling physiological features, and uses the disentangled features and noise to cross-generate the pseudo data. After that, the pseudo data are used to improve the robustness of the network. (3) The serious random noise and noisy data are difficult or even impossible to model explicitly. Thus, DRNet focuses on: how to obtain a physiological estimator that can extract noise-free features, instead of wasting on direct modeling noise or noisy data.
\begin{table}[t]
\footnotesize
\centering
\setlength\tabcolsep{5pt}
\caption{The ablation study of DRNet for HR estimation on the VIPL-HR database.}
\begin{tabular}{lrrrrrrr}
\hline
Method        & PC & SAB & Std$\downarrow$ & MAE$\downarrow$ & RMSE$\downarrow$  & MER$\downarrow$ & r$\uparrow$\\
\hline
Baseline      & $\times$     & $\times$ &  8.70 & 5.66 & 8.77 & 6.93\% & 0.74 \\
              & $\surd$      & $\times$ &  7.53 & 4.71 & 7.56 & 5.81\% & 0.81 \\
              & $\surd$      & $\surd$  &  7.18 & 4.47 & 7.29 & 5.45\% & 0.83 \\
DRNet         & $\surd$      & $\surd$  &  \textbf{6.75} & \textbf{4.18} & \textbf{6.78} & \textbf{5.14\%} & \textbf{0.85}   \\
\hline
\end{tabular}
\label{tab:ablation_vipl_hr}
\end{table}

%% file: 7_Conclusions.tex
\section{Conclusion}
In this paper, we propose an effective end-to-end network for remote physiological sensing using a decomposition and reconstruction strategy to reduce the influences of non-physiological signals and enhance feature disentanglement. Moreover, we design a lightweight and efficient Spatial Attention Block. Besides, a novel Patch Cropping augmentation strategy is proposed for enriching the training data. Extensive experiments are performed to verify the effectiveness of the proposed methods. In the future, we will explore the self-supervised learning technologies for remote physiological measurement.

%% file: 8_Acknowledgment.tex
\section*{Acknowledgments}
This work was supported in part by the NSFC-Xinjiang Joint Fund under Grant U1903127 and in part by the Natural Science Foundation of Shandong Province under Grant ZR2020MF052.